\documentclass[twocolumn,showpacs,superscriptaddress,preprintnumbers,prc]{revtex4-1}
\usepackage{graphicx,amsmath,amssymb,dcolumn,bm}
\usepackage{fixmath}
\usepackage{epic,eepic}
\usepackage{slashed}

\allowdisplaybreaks[2]
\usepackage{calrsfs}
\DeclareMathAlphabet{\pazocal}{OMS}{zplm}{m}{n}
\def\pt{\widetilde p}
\def\pa{p_A}
\def\pn{p_N}

\begin{document}
\title{Existence of nuclear modifications \\
of the nucleon longitudinal-transverse structure-function ratio}
\author{S. Kumano}
\affiliation
{Quark Matter Research Center,
    Institute of Modern Physics, Chinese Academy of Sciences,\\
    Lanzhou, 730000, China,\\
 Southern Center for Nuclear Science Theory,
    Institute of Modern Physics, Chinese Academy of Sciences,\\
    Huizhou, 516000, China}
\affiliation
{Department of Mathematics, Physics, and Computer Science,
    Faculty of Science, Japan Women's University,\\
    Mejirodai 2-8-1, Tokyo, 112-8681, Japan}
\affiliation
{KEK Theory Center, Institute of Particle and Nuclear Studies, KEK,
Oho 1-1, Tsukuba, 305-0801, Japan}
\date{November 3, 2025}

\begin{abstract}
It has been assumed that nuclear modification does not exist
in the longitudinal-transverse structure-function ratio 
$R_N=F_L^N/(2xF_1^N)$ in lepton deep inelastic scattering.
This assumption is widely used in obtaining structure functions 
of the ``nucleon" from nuclear data such as the deuteron ones.
However, nuclear modifications do exist theoretically
at least in medium- and large-$x$ regions
because nucleons in a nucleus move in any direction,
which is not necessarily the longitudinal direction 
of the virtual-photon or weak-boson momentum in lepton scattering.
Because of this transverse motion, the nucleon's transverse 
and longitudinal structure functions should mix with each other 
in nuclei with the mixture probability proportional 
to the nucleon's transverse momentum squared $\vec p_T^{\,\, 2}/Q^2$.
In this work, numerical results are explicitly shown 
regarding such nuclear modifications in the deuteron.
These nuclear modifications are important for determining
precise structure functions of the nucleon. Furthermore, 
modifications of $R_N$ should be investigated also 
at small $x$ by the future electron-ion collider
to find interesting gluon dynamics in nuclei.
Hopefully, this nuclear effect of $R_N$ could be found by future 
experimental measurements at lepton accelerator facilities.
\end{abstract}
\maketitle

\section{Introduction}
\label{intro}

Lepton deep inelastic scattering from the nucleon has been investigated
from the 1970's, and now the details have became clear 
about the internal structure of the nucleon
in the unpolarized parton distribution functions (PDFs). 
The polarized PDFs have been also determined and 
the polarized gluon part is expected to be determined 
by future electron-ion colliders (EICs) \cite{eic,lhec,eicc}.
In recent years, such studies have been extended to investigate
three dimensional structure functions in terms of
transverse-momentum-dependent parton distribution functions
\cite{tmds}, generalized parton distributions (GPDs) \cite{gpds}, 
and generalized distribution amplitudes
(or $s$-channel GPDs) \cite{gdas}.
Currently, these experimental studies are actively done mainly
at the Thomas Jefferson National Accelerator Facility (JLab) 
\cite{jlab}, and the projects will be continued at the EICs.

In charged-lepton deep inelastic scattering from the unpolarized 
nucleon, there are two structure functions $F_1^N$ and $F_2^N$.
Here, $N$ indicates the nucleon and it is explicitly written
in this paper to distinguish them from the nuclear ($A$) 
and deuteron ($D$) structure functions. 
The function $F_1^N$ is the structure function 
for the transverse polarization of the virtual photon, 
and the $F_2^N$ has components of both 
the transverse and longitudinal polarizations.
The longitudinal structure function $F_L^N$ is defined as
$F_L^N = (1+Q^2/\nu^2) F_2^N - 2 x F_1^N$ in terms of 
$F_1^N$ and $F_2^N$ with the four-momentum transfer $q$ ($Q^2 = -q^2$),
the energy transfer $\nu$, and the Bjorken scaling variable 
$x=Q^2/(2 M_N \nu)$, where $M_N$ is the nucleon mass.
The longitudinal-transverse structure function ratio $R$
is then given by $R_N=F_L^N/(2xF_1^N)$ \cite{quark-lepton-book}.

Nuclear modifications of the structure function $F_2^N$ are now
well known from small $x$ to large $x$ for many nuclei.
At small $x$, there are negative modifications due to 
nuclear shadowing. At medium $x$, nuclear binding and 
possible nucleon's internal modifications contribute. 
At large $x$, the nuclear modification increases
due to the nucleon's Fermi motion in a nucleus 
\cite{emc-effect}.

On the other hand, there is no experimental evidence 
on the nuclear modification of the ratio $R_N$
at this stage as reported in 2001 and 2002
\cite{hermes-2000,R-neutrino}.
The longitudinal-transverse ratios were measured also at JLab 
in 2007 \cite{R-Jlab-2007}; however, a significant nuclear effect 
was not clear by considering experimental errors.
Therefore, it is taken as grated in analyzing nuclear data, 
$e.\, g.$ for the deuteron,
that the nuclear modification does not exist for the ratio $R_N$.
in order to obtain the structure functions of the ``nucleon". 
This assumption is used, for example,
in obtaining the polarized structure function $g_1$ 
for the neutron from the measurements of the deuteron and $^3$He.
In the similar way, the nucleon's function $R_N$ is used
even for heavy nuclei in studying short-range correlations.

To be precise, this assumption is not right theoretically 
as it was proposed in 2003 \cite{ek-2003}.
Because a nucleon in a nucleus moves in any direction,
which is not necessarily the longitudinal direction 
along the virtual photon momentum, 
due to its Fermi motion, the transverse and longitudinal 
structure function could mix.
This topic could become interesting again in the near future 
because a new experimental study of $R_N$ \cite{R-Jlab-pro-R}
is scheduled to run in 2026 \cite{JP-Chen} 
and because the polarized deuteron experiment 
will start soon at JLab \cite{JLab-deuteron}. 
Therefore, it is worth shedding light on this nuclear modification 
topic by using a reliable theoretical model.
Although I fofus on the medium- and large-$x$ regions
in this work, the small-$x$ nuclear modification 
should also be and interesting topic because $F_L^A$
is related to gluon dynamics in nuclei as investigated
at HERA (Hadron-Elektron-Ringanlage) for the nucleon \cite{hera-fl}
and it will be investigated at the EICs \cite{eic,lhec,eicc}.

In this paper, I show numerically that such nuclear modifications
of $R_N$ exist in the deuteron.
These nuclear modifications need to be considered 
for obtaining precise structure functions of the ``nucleon" 
from nuclear data.
This article consists of the following. 
First, the theoretical formalism is shown
for the nuclear modifications of the longitudinal-transverse
ratio $R_N$. Then, numerical results are shown and they are summarized.

\vspace{0.30cm}
\section{Formalism}
\label{formalism}

The cross section of charged-lepton deep inelastic scattering from
a nucleon or a nucleus is given by the lepton tensor 
$L^{\mu \nu}$ and the hadron tensor $W^{A,N}_{\mu \nu}$ as
\begin{equation}
\frac{d \sigma}{dE_\ell^{\,\prime} d\Omega_l^{\,\prime}} =
            \frac{|\vec p_\ell^{\, \prime}|}{| \vec p_\ell |} 
            \, \frac{\alpha^2}{(q^2)^2}
            \, L^{\mu \nu} (p_\ell, q)
            \, W^{A,N}_{\mu \nu} (p_{A,N}, q) ,
\label{eqn:cross1}
\end{equation}
where $E_\ell^{\,\prime}$ and $\Omega_l^{\,\prime}$
are the final charged-lepton energy and solid angle,
$p_\ell$ and $p_\ell^{\, \prime}$ are initial
and final charged-lepton momenta,
$q$ is the momentum transfer $q=p_\ell - p_\ell^{\, \prime}$,
$p_{_A}$ and $p_{_N}$ are nuclear and nucleon momenta, 
respectively, and $\alpha$ is the fine structure constant of 
the quantum electrodynamics.
The hadron tensor of the charged-lepton scattering is given
in general by two structure functions $W_1^{A,N}$ and $W_2^{A,N}$ 
for the unpolarized nucleon and nuclei.

The hadron tensor for the polarized photon with the helicity $\lambda$
is given by \cite{quark-lepton-book}
\begin{align}
W^{A,N}_\lambda (p_{_{A,N}}, q) 
            = \varepsilon_\lambda^{\mu *} \varepsilon_\lambda^{\nu} 
                  W^{A,N}_{\mu\nu} (p_{_{A,N}}, q) ,
\label{eqn:W_lambda}
\end{align}
where $\varepsilon_\lambda^{\mu}$ is the photon polarization vector.
Then, the transverse and longitudinal structure functions 
$W^{A,N}_T$ and $W^{A,N}_L$ are defined by these helicity functions,
$W^{A,N}_{1}$, and $W^{A,N}_{2}$ as
\begin{align}
W^{A,N}_T & = 
    \frac{ W^{A,N}_{\lambda=+1} + W^{A,N}_{\lambda=-1} }{2}
     = W^{A,N}_{1} ,
\nonumber \\
W^{A,N}_L & = W^{A,N}_{\lambda=0}
     = \left ( 1 + \frac{\nu_{_{A,N}}^2}{Q^2} \right) W^{A,N}_2 - W^{A,N}_1 ,
\end{align}
where $\nu_N$ and $\nu_A$ are defined by
\begin{align}
\nu_A = \frac{ p_{_A} \cdot q }{\sqrt{p_{_A}^{\,2}}} = \nu , \ \ 
\nu_N = \frac{ p_{_N} \cdot q }{\sqrt{p_{_N}^{\,2}}} .
\end{align}
Instead of $W_{1,2,T,L}$, the structure functions $F_{1,2,T,L}$ 
are often used. 
They are related with each other by the relations:
\begin{align}
F_1^{A,N} & = F_T^{A,N}  = \sqrt{p_{_{A,N}}^2} \, W_1^{A,N} ,
\nonumber \\
F_2^{A,N} & = \frac{p_{_{A,N}} \cdot q}{\sqrt{p_{_{A,N}}^2}} \, W_2^{A,N} ,
\nonumber \\
F_L^{A,N} & = \left ( 1 + \frac{Q^2}{\nu_{_{A,N}}^2} \right ) F_2^{A,N} 
            - 2 x_{_{A,N}} F_1^{A,N} .
\end{align}
From these structure functions, the longitudinal-transverse 
structure function ratio is defied as
\begin{align}
R_{A,N} (x_A, Q^2) = \frac{F_L^{A,N} (x_{A,N}, Q^2)}
 {2 \, x_{A,N} F_1^{A,N} (x_{A,N}, Q^2)}  .
\label{eqn:RAN}
\end{align}
Here, the variables $x_{_{A}}$ and $x_{_{N}}$ are defined 
in terms of the Bjorken scaling variable $x$
and the momentum fraction $y$ as
\begin{alignat}{2}
x_A & = \frac{Q^2}{ 2 \, p_A \cdot q}, & \ \ 
x_N & = \frac{Q^2}{ 2 \, p_N \cdot q} = \frac{x}{y},  
\nonumber \\
x   & = \frac{Q^2}{2 \, M_N \nu}, & \ \ 
y   & = \frac{p_N \cdot q}{M_N \, \nu},
\end{alignat}
where $M_A$ is the nuclear mass.

A conventional way to theoretically describe 
the nuclear structure functions at medium- and large-$x$
is to use the convolution formalism.
Then, the nuclear hadron tensor is given by the nucleonic one
convoluted with the nucleon momentum distribution $S(\pn)$,
which is called the spectral function, as
\begin{align}
W^A_{\mu\nu} (\pa, q) 
  = {\displaystyle\int} d^4 \pn \, S(\pn) \, W^N_{\mu\nu} (\pn, q).
\end{align}
A simple form of the spectral function is given
by the shell model as
\begin{align}
\! \! \! \!
S (p_N) \! = \! \sum_i | \phi _i (\vec p_N) |^2  \delta \!
   \left ( p_N^{\, 0} - M_A + \sqrt{M_{A-i}^{\ 2} 
          +\vec p_N^{\ 2}} \, \right ).
\end{align}
Here, $\phi_i$ is the wave function of the nucleon $i$,
the separation energy $\varepsilon_i$ and $M_{A-i}$
are related by $\varepsilon_i = (M_{A-i}+M_N) - M_A$.
I average over all the nucleons 
for calculating the average separation energy, 
$\varepsilon_i \rightarrow \left< \varepsilon \right>$.
By using a non-relativistic approximation for
$\sqrt{M_{A-i}^{\ 2}+ \vec p_N^{\ 2}}$, 
$p_N^{\, 0}$ and $\left< \varepsilon \right>$ are related by 
\begin{align}
p_N^{\, 0} = M_N - \left< \varepsilon \right> 
   - \frac{\vec p_N^{\ 2}}{2 M_{A-1}},
\end{align}
with the replacement $M_{A-i} \to M_{A-1}$
\cite{hksw-2011,b1-conv}.
The spectral function contains the binding and Fermi-motion effects
by the change of the nucleon's effective mass.
Here, the nucleons are off-shell; however, the structure functions
of the on-shell nucleons are used.
Meson-exchange currents could also contribute, but they are
neglected in this work.

Applying the projection operators $\widehat P_{1,2}^{\, \mu\nu}$
to $W_{\mu\nu}^A$, we obtain the nuclear structure functions $W_{1,2}^A$ 
in the convolution model.
From $W_{1,2}^A$, the transverse and longitudinal structure functions
$F_1^A$ and $F_L^A$ are obtained in the convolution form as
\begin{align}
\! \! \! \! 
\left(
    \begin{aligned}
      \,      F_2^A(x_A,Q^2) \, \\
      \, 2x_A F_1^A(x_A,Q^2) \, \\
      \,      F_L^A(x_A,Q^2) \,
    \end{aligned}
\right)
& \! = \! \int_x^A \! dy
\left(
    \begin{aligned}
      \, f_{22} (y) \, & \ \ \ \, 0 \,      & \, 0 \ \ \ \,  \\
      \, 0 \ \ \ \     & \, f_{11} (y) \, & \, f_{1L} (y) \,\\
      \, 0 \ \ \ \    & \, f_{L1} (y) \, & \, f_{LL} (y) \,
    \end{aligned}
\right)
\nonumber \\
& \ \ \ \ \ \ \ \ \ 
\times
\left(
    \begin{aligned}
      \,               F_2^N(x/y,Q^2) \, \\
      \, 2 \, \frac{x}{y} F_1^N(x/y,Q^2) \, \\
      \,               F_L^N(x/y,Q^2) \,
    \end{aligned}
\right) .
\label{eqn:12L-convolution}
\end{align}
Here, the nucleon's lightcone momentum distributions
$f_{22}$, $f_{LL}$, $f_{L1}$, $f_{1L}$, and $f_{11}$
are defined by \cite{ek-2003}
\begin{align}
& f_{22} (y) 
= \int_0^\infty dp_{N\perp} 2 \pi p_{N\perp} y
             \frac{M_N \nu}{|\vec q \, |} 
             \left | \phi (\vec p_N) \right |^2
\nonumber \\
&  \ \hspace{1.3cm} 
\times
  \left [ \frac{2 \, (x/y)^2 \, \vec p_{N\perp}^{\ 2}}{(1+Q^2/\nu^2)Q^2}
 + \left ( 1 + \frac{2 \, (x/y) \, 
        p_{N\parallel}}{\sqrt{Q^2+\nu^2}} \right )^2 \right ],
\nonumber \\
& f_{LL} (y) = f_{11} (y) 
\nonumber \\
& \ \ \ \ \  
= \int_0^\infty dp_{N\perp} 2 \pi p_{N\perp} y
             \frac{M_N \nu}{|\vec q \, |} 
             \left | \phi (\vec p_N) \right |^2
    \left ( 1 + \frac{\vec p_{N\perp}^{\ 2}}{\pt_N^{\,\, 2}} \right ) ,
\nonumber \\
& f_{L1} (y) = f_{1L} (y)
\nonumber \\
& \ \ \ \ \  
= \int_0^\infty dp_{N\perp} 2 \pi p_{N\perp} y
             \frac{M_N \nu}{|\vec q \, |} 
             \left | \phi (\vec p_N) \right |^2
             \frac{\vec p_{N\perp}^{\ 2}}{\pt_N^{\,\, 2}} ,
\label{eqn:fL1}
\end{align}
where $\pt_{\mu} = p_{\mu} -(p \cdot q) \, q_\mu /q^2$.
Then, the longitudinal-transverse ratio $R_A$ of a nucleus is 
calculated by Eq.\,(\ref{eqn:RAN}).
This convolution model has been used for studying
the structure function $F_2^A$.
Here, I assume it is valid also for $F_1^A$ and $F_L^A$.
Higher-twist effects are not taken into account.
Here, $\pt_N^{\,\, 2}$ is expressed by $Q^2$ as 
\begin{align}
\! \! 
\pt_N^{\,\, 2} = \frac{Q^2}{4 x_N^2} \!
                 \left( 1+\frac{4 x_N^2 p_N^2}{Q^2} \right)
            \simeq \frac{Q^2}{4 x_N^2} \!
                 \left( 1+\frac{4 x_N^2 M_N^2}{Q^2} \right) .
\label{eqn:ptN2-Q2}
\end{align}

Because of the transverse motion of the nucleons in a nucleus,
the transverse and longitudinal structure functions of the nucleon
($F_1^N$, $F_L^N$) mix with each other 
in Eqs.\,(\ref{eqn:12L-convolution}) and (\ref{eqn:fL1}), 
and the mixture coefficient is proportional
to the transverse momentum squared $\vec p_{N\perp}^{\ 2}$.
The mixture factor $\vec p_{N\perp}^{\ 2}/\pt_N^{\,\, 2}$ is 
proportional to $\vec p_{N\perp}^{\ 2}/Q^2$.
Because of this admixture, I definitely
can claim that the nuclear modifications of $R_N$ do exist
although there is no experimental finding at this stage.
Because of the factor $\vec p_{N\perp}^{\ 2}/Q^2$,
this mixture does not occur in the Bjorken scaling limit 
$Q^2 \to \infty$.

However, one may note that the nuclear modifications
of $R_N$ exist even at $Q^2 \to \infty$ through 
the convolution integrals of Eq.\,(\ref{eqn:12L-convolution})
as shown numerically later. This modification occurs 
because the $x_N$-dependent functional forms
are different between $F_1^N$ and $F_L^N$, although
the lightcone momentum distributions of the nucleon,
$f_{LL}(y)$ and $f_{11}(y)$, are same.
In Eq.\,(\ref{eqn:12L-convolution}),
the structure function $F_2^A$ is expressed only by $F_2^N$
and the mixing with other structure function does not exist; 
however, there are correction factors $\vec p_{N\perp}^{\ 2}/Q^2$ 
and $p_{N\parallel} / \sqrt{Q^2+\nu^2}$ 
in $f_{22}(y)$ of Eq.\,(\ref{eqn:fL1}).
By the convolution description for the nuclear structure functions,
we found that there are two sources about the nuclear modifications
of $R_N$. One is the admixture of $F_1^N$ and $F_L^N$ 
in a nuclear medium, and the other is by the convolution 
with the nucleon's momentum distributions 
$f_{11} (y)$ and $f_{LL}(y)$.
In this work, the nuclear modifications are actually shown 
numerically for the deuteron by taking 
an appropriate deuteron wave function $\phi_i$
in Eq.\,(\ref{eqn:fL1}).

Next, I explain comparisons with related works.
There were works regarding convolution model calculations
on nuclear structure functions of light nuclei including the deuteron.
For example, the functions $F_2^A$ were calculated
for the deuteron, $^3$He, and $^3$H in Ref.\,\cite{pssk-2001}.
The functions $F_T^A$, $F_2^A$, and $F_3^A$ were investigated
also by including target-mass corrections (TMCs), off-shell effects,
and pion contributions in Ref.\,\cite{kp-2006}.
These nuclear binding, Fermi motion, TMCs, 
and off-shell effects are shown in detail in comparison
with the SLAC and JLab experimental data \cite{tems-2019}.
The nuclear-binding and Fermi-motion parts of these works
are essentially the same as the work of Ref.\,\cite{ek-2003}
and this work as shown in Eqs.\,(\ref{eqn:12L-convolution})
and (\ref{eqn:fL1}) by considering the factor 
in Eq.\,(\ref{eqn:ptN2-Q2}).
It is noteworthy that the nuclear modifications of $R_N$
are affected by the TMCs. In Ref.\,\cite{bahm-2011},
different prescriptions of the TMCs are shown numerically
by using the operator product expansion (OPE),
the collinear factorization of Ellis, Furmanski, and Petronzio (EFP),
$\xi$ scaling, and Accardi and Qiu (AQ) method.

\vspace{0.30cm}
\section{Results}
\label{results}
In order to calculate the nuclear modifications of $R_N$ by using
Eqs.\,(\ref{eqn:RAN}), (\ref{eqn:12L-convolution}),
and (\ref{eqn:fL1}), I need to supply the information 
on the used functions.
First, the structure function of the nucleon $F_2^N$ is calculated
by the PDFs of MSTW08 \cite{mstw08}
in the leading-order of the running coupling constant $\alpha_s$. 
The function $R_N$ of the nucleon is given by
the SLAC parametrization of 1990 \cite{r1990}.
Using these functions, $F_1^N$ is calculated by
$F_1^N =  ( 1+Q^2/\nu^2 ) \, F_2^N / [2 \, x \, (1+R_N)]$.
As for the wave function $\phi$ of the deuteron,
the Bonn wave function \cite{deuteron} is used.

\begin{figure}[b]
\vspace{-0.30cm}
\begin{center}
   \includegraphics[width=8.5cm]{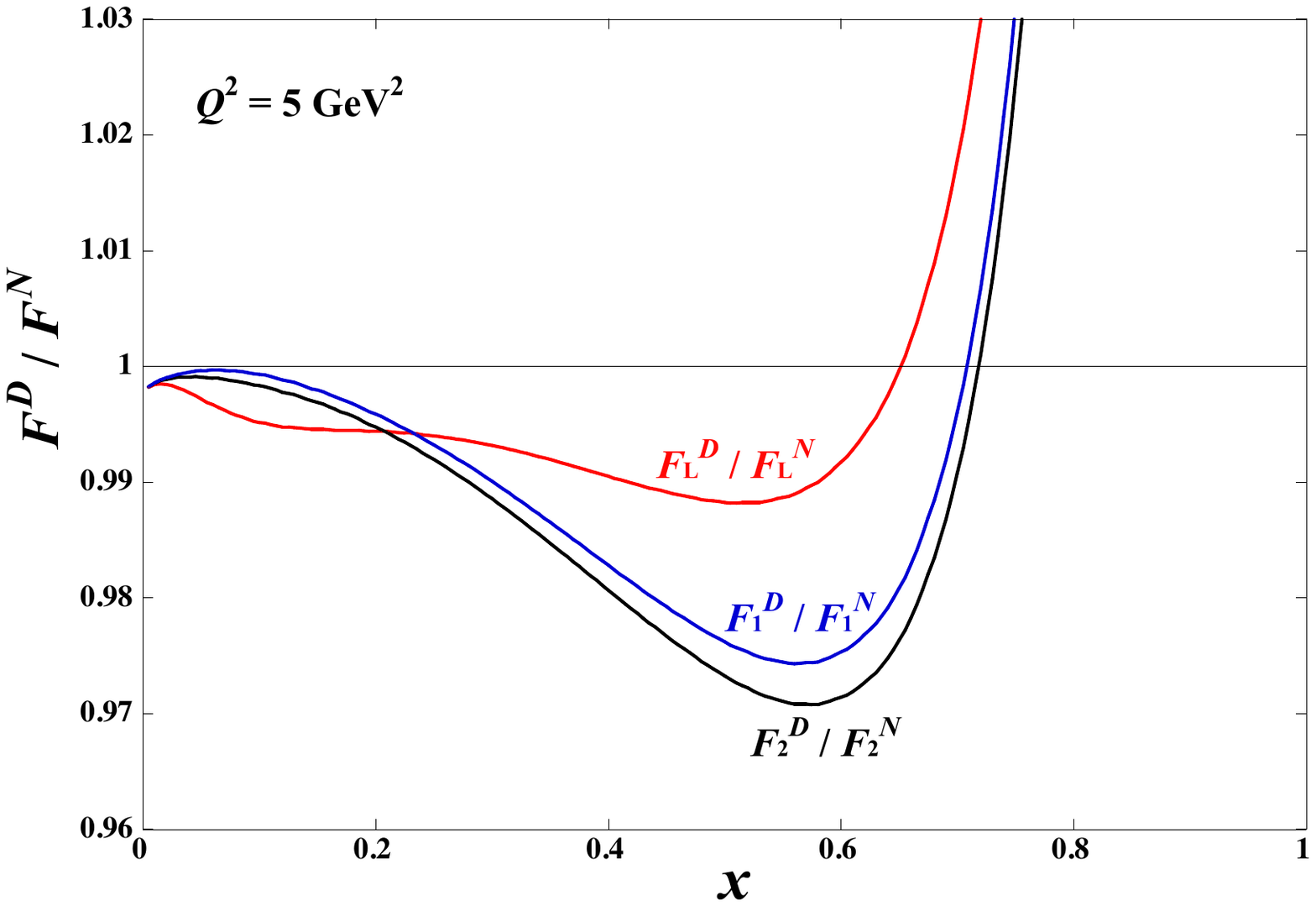}
\end{center}
\vspace{-0.9cm}
\caption{Nuclear modifications of the structure functions
$F_1$, $F_2$, and $F_L$ at $Q^2 =5$ GeV$^2$ in the deuteron.}
\label{fig:F2F1FL}
\vspace{-0.00cm}
\end{figure}

With these preparations, the nuclear structure functions
$F_2^A$, $F_1^A$, and $F_L^A$ are calculated, and 
their nuclear modifications are shown in Fig.\,\ref{fig:F2F1FL}
at $Q^2=5$ GeV$^2$. 
The nuclear modifications of the transverse structure function
($F_1^D/F_1^N$) are very similar to the ones of 
$F_2^D/F_2^N$.
However, the modifications of the longitudinal structure function
($F_L^D/F_L^N$) are very different. Especially, the modification
dip in the medium $x$ region ($x \sim 0.5$) is not large.
This significant difference comes from the mixture effects
$\vec p_{N\perp}^{\ 2} / \pt_N^{\,\, 2}$ 
($\sim \vec p_{N\perp}^{\ 2} / Q^2$) in Eq.\,(\ref{eqn:fL1}).
In fact, if these mixing terms are terminated, the longitudinal
modification curve $F_L^D/F_L^N$ is very similar 
to the ones of the transverse and $F_2$ ones
($F_1^D/F_1^N$, $F_2^D/F_2^N$).
Because the nuclear modifications of $F_L^N$ are so different
from the ones of $F_2^N$, one should not assume that 
they are same in handling nuclear data which contain $F_L^A$.

\begin{figure}[t]
 \vspace{-0.00cm}
\begin{center}
   \includegraphics[width=8.5cm]{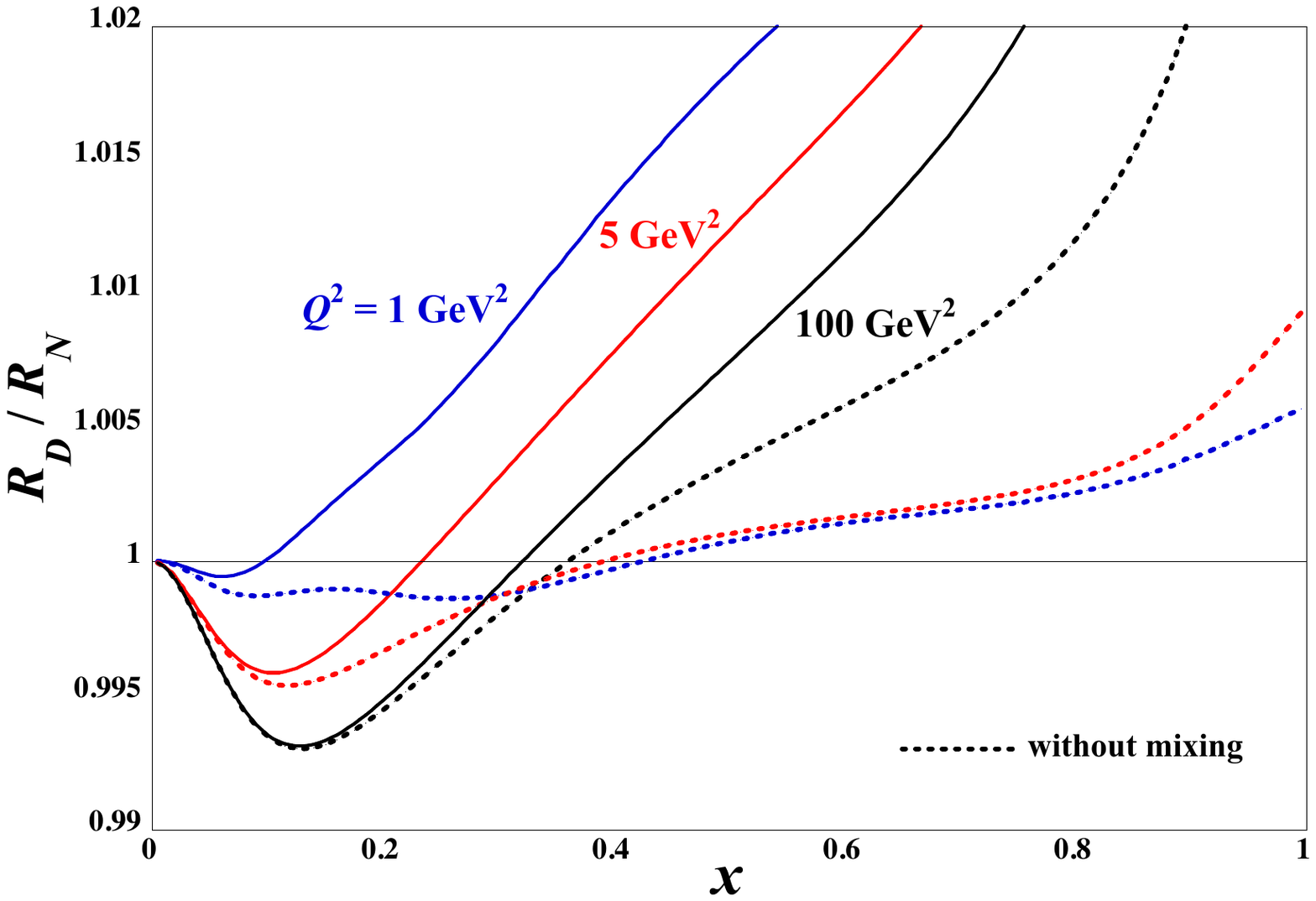}
\end{center}
\vspace{-0.9cm}
\caption{Nuclear modifications of the longitudinal-transverse 
structure function ratio $R_N$ at $Q^2 =1$, 5, and 100 GeV$^2$ 
in the deuteron. The dotted curves are obtained by terminating
the mixing terms.}
\label{fig:R-modification}
\vspace{-0.30cm}
\end{figure}

Next, I show the nuclear modifications in the form of 
the longitudinal-transverse ratio $R_A/R_N$ by 
using Eq.\,(\ref{eqn:RAN}) in Fig.\,\ref{fig:R-modification}
at $Q^2=1$, 5, and 100 GeV$^2$. The modifications
are shown by the solid curves, and the dotted ones
are obtained by terminating the mixing terms.
It means that 
$\vec p_{N\perp}^{\ 2} / \pt_N^{\,\, 2} \to 0$ is taken 
in the expressions of $f_{LL} (y)$, $f_{11} (y)$, $f_{L1} (y)$, 
and $f_{1L} (y)$ of Eq.\,(\ref{eqn:fL1})
and that the replacement $\big [ \cdots \big ] \to 1$
is taken in $f_{22} (y)$.
Because the major source of the nuclear modifications of $R_N$
is the longitudinal-transverse mixing
($\sim  \vec p_{N\perp}^{\ 2} / Q^2$),
the modifications are large at small $Q^2$ ($\sim 1$ GeV$^2$).
As the $Q^2$ increases as $Q^2=1 \to 5 \to 100$ GeV$^2$,
these modifications becomes smaller.
However, the significant modifications still exist even at large $Q^2$.
In fact, we notice that the modifications are noticeable
at $Q^2=100$ GeV$^2$ even without the mixing terms 
as shown by the dotted curve of $Q^2=100$ GeV$^2$, 
and such modifications are obtained
through the convolution with the nucleon's lightcone 
momentum distribution $f_{LL} (y)$.
At smaller $Q^2$ ($=1,\ 5$ GeV$^2$), the nuclear modifications
in $R_N$ are not large if there is no mixing effect
as shown by the dotted curves.
One should note that the function $R_N$ is not accurately 
determined especially at large $x$, so that 
the current numerical results at large $x$ 
could have large uncertainties at this stage. 

The nucleon's PDFs and structure functions have been determined
by partially using nuclear data including the deuteron ones.
In fact, the deuteron is often used for determining
the nucleon's, especially neutron's, structure functions 
by assuming that the nuclear modification does not exist 
in the longitudinal-transverse structure function ratio $R_N$. 
However, it is obviously not appropriate according to 
the results shown in this work.
As shown in Fig.\,\ref{fig:F2F1FL}, the nuclear modifications
of the deuteron's structure functions are of the order 
of a few percent. The modifications of $R_N$ are 
of similar magnitude or less in Fig.\,\ref{fig:R-modification}.
The effects may be larger at large $x$; however,
it is difficult to predict the large-$x$ 
region of $R_D$ due to the possible uncertainty of $R_N$ itself. 
Obviously experimental efforts are needed 
to accurately measure not only the functions $R_A$ 
for various nuclei but also the $R_N$ for the nucleon
including the large-$x$ region.

In comparison with the actual experimental data,
one should note that the TMCs could affect the modification curves
of Fig.\,\ref{fig:R-modification} especially at small $Q^2$.
Modifications of the nucleon structure functions 
by using the different prescriptions of the TMCs, namely
the OPE, EFP, $\xi$ scaling, and AQ methods
at $Q^2=2$ GeV$^2$ \cite{bahm-2011}.
For the deuteron, the large Fermi-motion effects are 
reduced by these TMCs depending on the prescription.
If the TMCs are used for both deuteron and nucleon
structure functions, their effect partially cancel
with each other in taking the deuteron/nucleon ratios.
Therefore, the carful and detailed comparisons 
should be made with the data in future by including 
the TMCs especially for comparison with the JLab deuteron data
in the near future \cite{R-Jlab-pro-R,JP-Chen,JLab-deuteron}.

\begin{figure}[t]
 \vspace{-0.00cm}
\begin{center}
   \includegraphics[width=8.5cm]{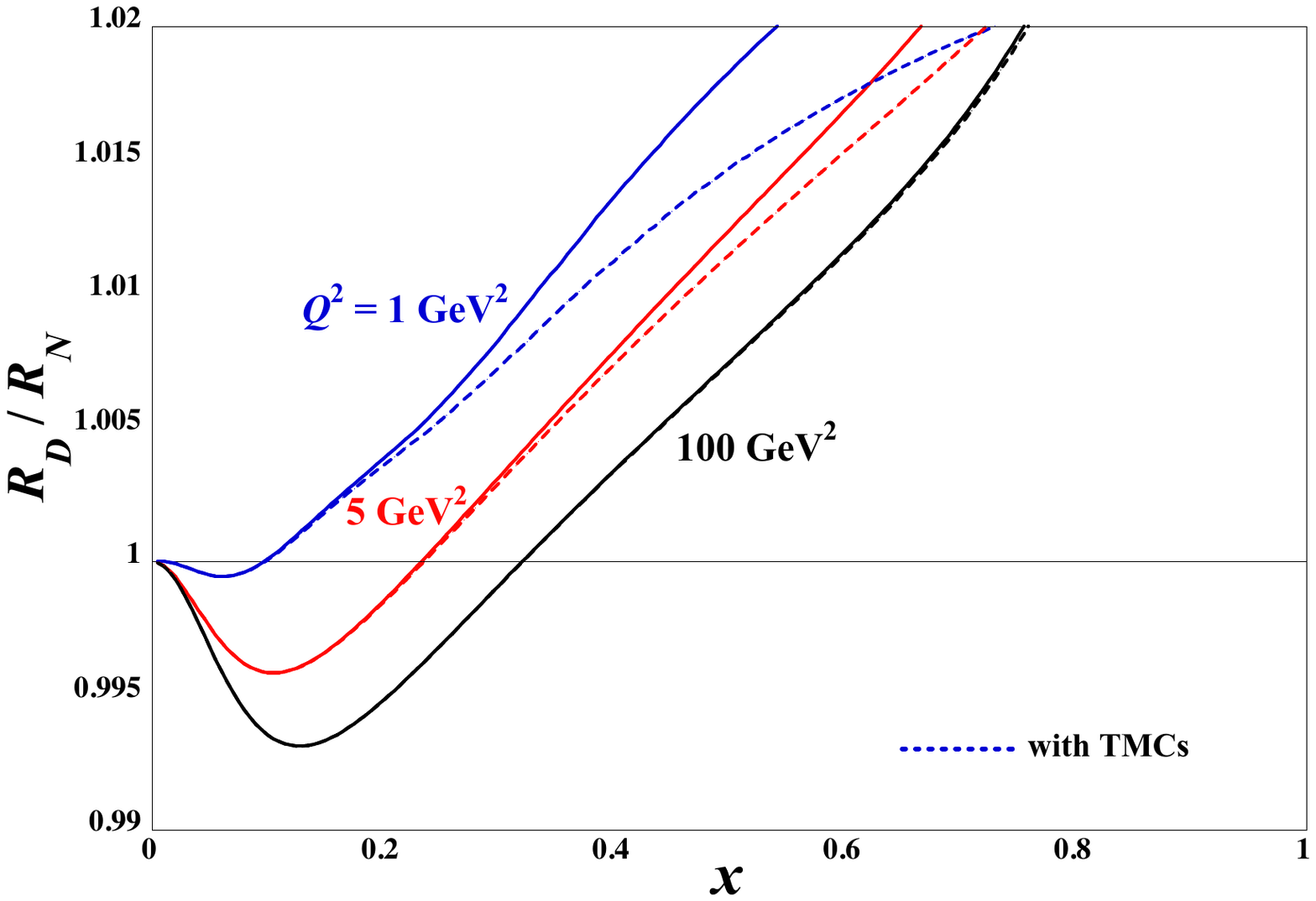}
\end{center}
\vspace{-0.9cm}
\caption{Nuclear modifications of the longitudinal-transverse 
structure function ratio $R_N$ at $Q^2 =1$, 5, and 100 GeV$^2$ 
in the deuteron. The dashed curves are the modifications
with the TMCs given by the $\xi$ scaling.}
\label{fig:R-modification-TMCs}
\vspace{-0.30cm}
\end{figure}

As mentioned in the end of Sec.\,\ref{formalism},
there are various prescriptions of the TMCs, for example,
the OPE, EFP, $\xi$ scaling, and AQ methods.
According to Ref.\,\cite{bahm-2011},
the $\xi$ scaling and the AQ prescriptions suppress
the Fermi-motion rise in the nuclear structure functions
at large $x$, and the OPE and EFP methods suppress
more significantly.
In order to illustrate such effect, numerical results are
shown by taking the $\xi$ scaling as an example.
The $\xi$ scaling formulas are given in Ref.\,\cite{bahm-2011},
where one may note that the TMCs have different overall
factors among $F_1^{A,N}$, $F_L^{A,N}$, and $F_2^{A,N}$.
The nuclear modifications of $R_N$ are also changed.
The TMCs of the $\xi$ scaling are shown
by the dashed curves for the nuclear modification $R_A/R_N$ 
in Fig.\,\ref{fig:R-modification-TMCs}.
As shown in Ref.\,\cite{bahm-2011}, the TMCs suppress
the Fermi-motion rise and such effects are large 
at small $Q^2$ ($=1$ GeV$^2$) because of 
the $M_N^2/Q^2$ corrections.
Since the TMCs depend greatly on the used prescriptions,
careful analyses should be done in future in comparison with
the deuteron data.


This kind of study could affect the polarized PDFs of the nucleon. 
As the longitudinally-polarized PDFs become accurate,
we need to be careful about the modifications of $R_N$
in nuclei because the deuteron and $^3$He data are used
for extracting the polarized PDFs without such modifications.
In the longitudinally-polarized charged-lepton nucleon 
deep inelastic scattering, experimental data are shown
by the spin asymmetry $A_1$. It is expressed as
$A_1 \simeq g_1^N /F_1^N = g_1^N \cdot 2x_N (1+R_N) / [ (1+Q^2/\nu_N^2) F_2^N] $
in term of the function ratio $R_N$. 
If the target is the polarized deuteron or $^3$He,
one needs to use $R_D$ or $R_{^3 \rm{He}}$.
However, in the current global analyses, it has been assumed
that there is no nuclear modification, namely 
$R_D = R_{^3 \rm{He}} = R_N$.
This assumption is not appropriate,
as the polarized PDFs become accurate, 
by considering the current research results of $R_D$
in Fig.\,\ref{fig:R-modification}.
Needless to say, if the target is a larger nucleus,
the nuclear modifications become more significant
and they should be taken into account.
In order to obtain accurate unpolarized and polarized structure
functions of the nucleon and then to apply to other high-energy
hadron reactions, it is important to consider the nuclear 
modifications of $R_N$.
The predicted nuclear modifications of $R_N$ could be measured
at JLab \cite{jlab,R-Jlab-pro-R} and 
at the future electron-ion colliders \cite{eic,lhec,eicc}.

One of the recent major achievements of JLab is
the finding of short-range correlations (SRCs) in nucleon-nucleon
interactions by using electron-nucleus scattering \cite{SRC-JLab}.
In the SRC measurements, used nuclei range from the deuteron as
the smallest nucleus to a large nucleus of lead ($^{208}$Pb).
The electron-nucleus cross section is generally expressed by $F_1^A$
and $F_2^A$, although the experimental kinematical region is
different from the deep inelastic one.
Obviously, nuclear modifications should be large 
in the longitudinal-transverse ratio $R_A$; however, 
they are neglected in the experimental analysis of the SRCs 
and simply the nucleon's ratio $R_N$ has been used \cite{SRC-Eli}. 
Therefore, the nuclear-modification issue should be readdressed
for the longitudinal-transverse ratio by throwing away 
the preconception that such a modification does not exist.

Although I focused the theoretical studies in this work
at medium and large $x$ for discussing the nuclear modification
effects from the nuclear binding and Fermi motion,
the nuclear effect of $R_N$ should be interesting 
also at small $x$. In fact, the $F_L$ data of HERA at small $x$ 
is related to the gluon distribution
in the nucleon \cite{hera-fl}.
Therefore, the nuclear modification of $R_N$ is 
an interesting quantity to investigate 
gluon dynamics in nuclei.
The studies regarding the nuclear modifications of $R_N$
contain not only the standard nuclear physics,
in terms of nuclear binding and Fermi motion,
but also possibly the gluon saturation physics in nuclei.
Therefore, such studies could lead to another interesting
new project at future EICs because the nuclear gluon dynamics
is an important aspect.

\vspace{0.30cm}
\section{Summary}
\label{summary}

Using the standard convolution description for the structure functions
of the deuteron, I calculated nuclear modifications of 
$F_2^N$, $F_1^N$, and $F_L^N$.
Because the nucleons in the deuteron move in any direction
which is generally different from the longitudinal direction
defined by the virtual photon momentum in the charged-lepton
scattering, the longitudinal structure function $F_L^N$
and the transverse one $F_1^N$ mix with each other in the deuteron.
Because of this mixture, the nuclear modification of $F_L^N$ is very
different from the ones of $F_2^N$ and $F_1^N$.
In addition, the modification of $R_N$ comes from
the convolution integrals with the nucleon's momentum distributions.
These mechanisms led to significant nuclear modifications of $R_N$
in the deuteron. It has been assumed until now that the nuclear
modification does not exist in the function $R_N$; therefore, 
our theoretical predictions should be tested by future experiments.

\vspace{0.30cm}
\begin{acknowledgements}
I was partially supported by a research program 
of Chinese Academy of Sciences and 
Japan Society for the Promotion of Science (JSPS) Grants-in-Aid
for Scientific Research (KAKENHI) Grant Number 24K07026.
I thank Jian-Ping Chen for suggestions about JLab experiments,
Eliezer Piasetzky for short-range correlation analysis,
and S. Hiyoshi for discussions.
\end{acknowledgements}



\end{document}